\documentclass[a4,11pt]{article}
\usepackage{amsmath,amsthm,amssymb,braket,bm}
\usepackage{comment}
\usepackage{jcappub}

\title{Fermionic Schwinger effect and induced current in de Sitter space}
\author[a, b]{Takahiro Hayashinaka, }
\author[c]{Tomohiro Fujita}
\author[a, b, d]{Jun'ichi Yokoyama}
\affiliation[a]{Research Center for the Early Universe (RESCEU), Graduate School of Science, 
The University of Tokyo, \\
Bunkyo-ku, Tokyo, 113-0033, Japan}
\affiliation[b]{Department of Physics, Graduate School of Science, 
The University of Tokyo, \\
Bunkyo-ku, Tokyo, 113-0033, Japan}
\affiliation[c]{Stanford Institute for Theoretical Physics and Department of Physics, 
Stanford University, Stanford, \\
CA 94306, USA}
\affiliation[d]{Kavli Institute for the Physics and Mathematics of the Universe (Kavli IPMU), 
WPI, UTIAS, The University of Tokyo, \\
Kashiwa, Chiba, 277-8583, Japan}

\date{\today}
\emailAdd{hayashinaka@resceu.s.u-tokyo.ac.jp}
\emailAdd{tomofuji@stanford.edu}
\emailAdd{yokoyama@resceu.s.u-tokyo.ac.jp}

\abstract{
We explore Schwinger effect of spin $1/2$ charged particles with static electric field in 1+3 dimensional de Sitter spacetime.
We analytically calculate the vacuum expectation value of the spinor current which is induced by the produced particles in the electric field. 
The renormalization is performed with the adiabatic subtraction scheme. 
We find that the current becomes negative, namely it flows in the direction opposite to the electric field, 
if the electric field is weaker than a certain threshold value depending on the fermion mass, which is also known to happen 
in the case of scalar charged particles in 1+3 de Sitter spacetime. 
Contrary to the scalar case, however, the IR hyperconductivity is absent in the spinor case. 
}

\keywords{Schwinger effect, de Sitter spacetime}

\arxivnumber{1603.04165}

\begin{document}

\maketitle

\section{Introduction}
The Schwinger effect \cite{schwinger1951gauge} is known as an intriguing example of non-perturbative phenomena 
of quantum field theory in a background field. 
It describes production of pairs of charged particles out of the quantum vacuum state 
due to a background electric field. 
It is still difficult to provide the electric field strong enough to cause the particle production due to Schwinger effect in a laboratory. 
However, we expect that such a strong electric field were naturally present in the primordial universe 
in the context of inflationary magnetogenesis 
which attempts to explain the origin of the observationally inferred large-scale magnetic fields 
\cite{neronov2010evidence, 2041-8205-747-1-L14, 2041-8205-744-1-L7, Schnitzeler01112010, Beck2011} 
in terms of the primordial magnetic fields generated during inflation~\cite{PhysRevD.37.2743}.
\footnote{For the review of magnetogenesis, see \cite{Kandus20111, Subramanian:2015lua}}

The observations of galactic and extra-galactic magnetic fields motivate the study of the origin of these magnetic fields 
and inflationary magnetogenesis is considered as a promising candidate 
as a way to achieve long enough coherent length. 
For instance, the kinetic coupling model (or $f^2 FF$ model \cite{ratra1992cosmological, PhysRevD.69.043507, PhysRevD.70.083508, martin2008generation}), 
which is a well-studied model of inflationary magnetogenesis, predicts that very strong electric fields are inevitably produced 
during inflation, if it generates magnetic fields which are strong enough to leave observable signatures. 
It is also known that the model cannot explain the lower bound on the present magnetic field strength inferred by the blazar observation, 
because of the overproduction of the electric fields whose energy density spoils the inflationary background \cite{demozzi2009magnetic} 
or the observation of the cosmic microwave background radiation \cite{1475-7516-2013-09-009}.%
\footnote{See however the recent works with extended models \cite{1475-7516-2014-05-040, Domenech:2015zzi, Fujita:2016qab}}
Nevertheless, no one has yet studied the model properly taking into account Schwinger effect, while it possibly changes the dynamics drastically. 
Therefore the Schwinger pair production could occur in the inflationary era and it is important to investigate it. 

Regarding Schwinger effect in de Sitter spacetime, two nontrivial backgrounds are involved. 
These are electric field and gravitational field, both of which can cause the particle production from the vacuum state. 
This combination of the different production sources makes the problem challenging and interesting. 
Furthermore, the background fields are not completely static in realistic situations and the backreaction to the electric field 
is important especially in the context of magnetogenesis. 
Flowing through the electric field, the produced particles induce the electric current and it affects the background electric field. 
The induced electric current which characterizes the size of the backreaction can be quantitatively evaluated by 
the vacuum expectation value of the current operator. 
Note that the induced electric current is a well-defined physical quantity since it only requires to specify the in-vacuum state, 
while the Bogoliubov coefficients which can give the particle production rate depend also on the definition of the out-vacuum state.

Recently, a number of studies of this effect in de Sitter spacetime have appeared 
\cite{PhysRevD.49.6343,Martin:2007bw,1475-7516-2014-04-009,Cai:2014qba,kobayashi2014schwinger,PhysRevD.93.025004}. 
Their motivations diverge into many branches of quantum physics 
from false vacuum decay and bubble nucleation to a thermal interpretation of particle production or cosmological consequences including magnetogenesis. 
Schwinger effect and its induced current in de Sitter spacetime for a \emph{scalar} charged particle have been investigated both in 
the $1+1$ dimension case~\cite{1475-7516-2014-04-009}, in the $1+2$ dimension case~\cite{Bavarsad:2016cxh} and the $1+3$ case~\cite{kobayashi2014schwinger}.
In those works, it is found that the scalar induced current is strongly enhanced for the small mass field and weak electric field regime. 
This phenomenon was called IR hyperconductivity and found in \cite{1475-7516-2014-04-009} for the first time. 
In \cite{kobayashi2014schwinger}, the authors reported a negative current which flows in a direction against the electric field 
in addition to the IR hyperconductivity. 
In the $1+3$ dimensional case, it was also found the terms which are not suppressed by the exponential factor 
$\exp(-\pi m^2/eE)$ or $\exp(-2\pi m/H)$ appear in the massive field limit. 
These suppression factors are naturally expected from the semiclassical approximation. Thus it suggests the breakdown of the 
semiclassical description in the massive limit. 
For \emph{spinor} charged particle, however, 
the induced current has been calculated only in the case of $1+1$ dimensional de Sitter spacetime \cite{PhysRevD.93.025004}.
In \cite{PhysRevD.93.025004}, the authors have shown that there is neither the IR hyperconductivity nor the negativity of the current.

To calculate the induced electric current in the real space, one has to address the divergence coming from vacuum contribution. 
In \cite{kobayashi2014schwinger, PhysRevD.93.025004}, the adiabatic subtraction method \cite{PhysRevD.9.341,PhysRevD.36.2963} was 
employed to remove the divergences in the vacuum expectation value of the induced current. 
It is known that the WKB (adiabatic) expansion for a fermionic field cannot 
satisfy the equation of motion while satisfying the normalization condition to all orders of the expanding parameter $\hbar$. 
Nevertheless, the equivalence of the adiabatic and the DeWitt-Schwinger renormalization schemes was shown in \cite{PhysRevD.91.064031}, 
and there are applications of the adiabatic regularization of fermionic fields recently \cite{PhysRevD.89.044030, PhysRevD.91.124075}. 

In this paper, we calculate 
the induced current for spinor QED in $1+3$ dimensional de Sitter space. 
We consider a static background electric field whose energy density is constant even in expanding de Sitter spacetime and employ the adiabatic subtraction as the regularization method. 
We analytically obtain the vacuum expectation value of the fermionic current and compare it with the scalar particle case and the semiclassical approximation \cite{:/content/aip/proceeding/aipcp/10.1063/1.4937198}. 
Furthermore, its weak/strong electric field limit and large mass limit are investigated in depth 
with the curious features found in the previous works in mind, such as the IR hyperconductivity, 
the negative current and the terms without the exponential mass suppression. 
We also discuss the stability of the background electric field considering the backreaction effect indicated by the electric current.

The rest of the paper is organized as follows. 
In Sec.~\ref{Sec:setups}, we introduce the treatment of the Dirac spinor in curved spacetime with clarifying our notation 
and solve the Dirac equation in a background electric field. 
In Sec.~\ref{Sec:inducedcurrent}, the calculation of the induced current is described. 
The property of the renormalized current is investigated in Sec.~\ref{Sec:result}. 
Finally, Section~\ref{Sec:conclusion} is devoted to the conclusion. 
Technical details can be found in appendices. 

\section{Setups}\label{Sec:setups}

\subsection{QED action in curved spacetime}
We start with the action for spinor quantum electrodynamics (QED) in curved spacetime, 
\begin{equation}\label{originalaction}
S = \int \mathrm d^4 x \sqrt{-g} \left\{ -\frac{1}{4}g^{\mu\alpha}g^{\nu\beta}F_{\mu\nu}F_{\alpha\beta} 
+ \bar \psi (i \gamma^\mu D_\mu - m) \psi \right\}, 
\end{equation}
where the metric sign is chosen as $(-+++)$. 
The covariant derivative $D_\mu$ is written in terms of U(1) gauge field $A_\mu$ and spinor connection $\Gamma_\mu$ as 
\begin{equation}
D_\mu(x) \equiv \partial_\mu + ieA_\mu(x) + \Gamma_\mu(x), 
\end{equation}
which ensures the local gauge symmetry and covariance. 
$F_{\mu\nu}=\nabla_\mu A_\nu- \nabla_\nu A_\mu$ is the gauge field strength. 

Introducing the tetrads $e_a^\mu$ ($a,b,c\cdots = 0,1,2,3$ are indices for the Local Lorentz Transformation (LLT) and 
the Greek indices $\mu, \nu\cdots$ are for spacetime) 
and the generators $\Sigma_{ab}$ of the LLT which obey the algebra 
$[\Sigma^{ab}, \Sigma^{cd}] = (\eta^{ac}\Sigma^{db}-\eta^{bc}\Sigma^{da}) - (\eta^{ad}\Sigma^{cb}-\eta^{bd}\Sigma^{ca})$, 
one can write down the spinor connection as 
\begin{equation}
\Gamma_\mu = \frac{1}{2} e_a^\nu e_{b\nu;\mu}\Sigma^{ab}. 
\end{equation}
$\Gamma_\mu$ then satisfies 
the correct transformation rule under the infinitesimal LLT $\Lambda_a^{\;b} = \delta_a^{\;b} + \omega_a^{\;b}$,
\begin{equation}\label{Gammatrns}\begin{split}
\Gamma_\mu(x)\rightarrow \bar\Gamma_\mu &= \frac{1}{2} \Lambda_a^{\;c} e_c^{\;\nu} (\Lambda_b^{\;d} e_{d\nu})_{;\mu}\Sigma^{ab}\\
&= \Gamma_\mu + \frac{1}{2}\omega_{ab}[\Sigma^{ab},\Gamma_\mu] - \omega_{ab,\mu}\Sigma^{ab}. 
\end{split}\end{equation}
Let $D(\Lambda) = 1+ \frac{1}{2}\omega_{ab}\Sigma^{ab}$ be a representation of the infinitesimal LLT, 
then Eq.~\eqref{Gammatrns} is rewritten as 
\begin{equation}
\bar\Gamma_\mu = D(\Lambda)\Gamma_\mu D^{-1}(\Lambda) - (\partial_\mu D(\Lambda))D^{-1}(\Lambda), 
\end{equation}
which gives a transformation nature 
\begin{equation}
e_a^{\;\mu}(\partial_\mu + \Gamma_\mu)\psi \rightarrow \Lambda_a^{\;b} e_b^{\;\nu}D(\Lambda)((\partial_\mu + \Gamma_\mu)\psi), 
\end{equation}
as required. 

For the Dirac spinor $\psi$, which is the reducible $(\frac{1}{2},0)\oplus(0,\frac{1}{2})$ representation of the LLT, 
the generators are given by $\Sigma^{ab} = -\frac{1}{4}[\gamma^a,\gamma^b]$. Here, the gamma matrices satisfy the anti-commutation relation 
$\{\gamma^a,\gamma^b\} = -2\eta^{ab}$ with $\eta^{ab}=\mathrm{diag} (-1,1,1,1)$ being the Minkowski metric. 

In the spatially flat Friedman-Lema{\^i}tre-Robertson-Walker spacetime $\mathrm ds^2 = a(\eta)^2(-\mathrm d\eta^2 + \mathrm d\bm{x}^2)$, 
the action Eq.~(\ref{originalaction}) reduces to
\begin{equation}
S = \int \mathrm d^4x\left\{ -\frac{1}{4}\eta^{\mu\alpha}\eta^{\nu\beta}F_{\mu\nu}F_{\alpha\beta} 
+ \bar\xi \left( i \gamma^a\partial_a - eA_a\gamma^a - ma \right)\xi\right\}, 
\end{equation}
where $\xi(\eta,\bm x)$ is the canonical Dirac field $\xi = a^{3/2}\psi$, and we have used the following equations hold in the FLRW metric: 
\begin{equation}
e_a^{\;\mu} =\frac{1}{a}\delta_a^{\;\mu},\;\;
\gamma^a e_a^{\;\mu}\Gamma_\mu = \frac{3a^\prime}{2a}\gamma^0,\;\; \gamma^a e_a^{\;\mu}(eA_\mu) = \frac{1}{a}eA_a\gamma^a, 
\end{equation}
where a prime denotes the derivative with respect to the conformal time $\eta$. 
It is clearly shown that the conformal symmetry is recovered for massless fermion case, $m=0$.

\subsection{Dirac equation in EM background}
The equation of motion for $\xi$ field (the Dirac equation) is given by 
\begin{equation}\label{Dirac1st}
(i \gamma^a\partial_a - eA_a\gamma^a - ma)\xi(\eta,\bm x)=0. 
\end{equation}
Substituting $\xi = (i \gamma^a\partial_a - eA_a\gamma^a + ma)\zeta$ into Eq.~(\ref{Dirac1st}), we obtain the quadratic Dirac equation, 
\begin{equation}\label{Dirac2nd}
\left\{(\partial_a+ieA_a)^2 -m^2 a^2 + i(ma^\prime\gamma^0 - \frac{e}{2}\gamma^a\gamma^bF_{ab})\right\}\zeta(\eta,\bm x)=0, 
\end{equation}
where all the indices should be contracted by the Minkowski metric $\eta^{ab}$. 

Hereafter we consider that a homogeneous electric background field $A_\mu(x)=(0,0,0,A_z(\eta))$ 
exists in de Sitter spacetime (i.e. $a^\prime/a^2=H=cosnt.$). 
We also assume that the background field $A_z$ is given by 
\begin{equation}\label{backgroundAz}
A_z = -\frac{E}{H}(a-1) = -\frac{E}{H}\left(\frac{1}{1-H\eta}-1\right), 
\end{equation}
where $E$ is a constant, the scale factor is taken as $a = 1/(1-H\eta)$ and the offset $-1$ is introduced in $(a-1)$ 
so that we can take Minkowski limit ($H\to0$) explicitly 
\begin{equation}
A_z \xrightarrow{H\to0} -Et. 
\end{equation}
Note that $\eta = (1-\mathrm e^{-H t})/H \xrightarrow{H\to0}t$, and $\eta \in (-\infty,1/H)$ in our notation. 
The physical strength of electric field in de Sitter spacetime is given by $-a^{-2} \partial_{\eta} A_z = E$, which yields a constant electromagnetic energy density. 

Let us further manipulate Eq.~\eqref{Dirac2nd}. It can be shown that 
$i(ma^\prime\gamma^0 - \frac{e}{2}\gamma^a\gamma^bF_{ab}) = iH^2 a^2 (\frac{m}{H}\gamma^0 + \frac{eE}{H^2}\gamma^0\gamma^3)$ and
\begin{equation}
\begin{split}
 \left(\frac{m}{H}\gamma^0 + \frac{eE}{H^2}\gamma^0\gamma^3\right)^2 =
 (M\gamma^0 + L\gamma^0\gamma^3)^2 
= (M^2+L^2)\bm 1, 
\end{split}
\end{equation}
where we have introduced two dimensionless parameters, 
\begin{equation}
M\equiv \frac{m}{H},
\qquad
L\equiv \frac{eE}{H^2}.
\end{equation}
which are the mass and the electric field strength normalized by the Hubble parameter. 
We find there exist four time-independent eigenvectors $w_s\;(s=1,2,3,4)$ of a matrix 
$B \equiv (M^2+L^2)^{-1/2} (M\gamma^0+L\gamma^0\gamma^3),$ 
\begin{equation}\label{defws}
B w_s = \lambda_s w_s, 
\end{equation}
which have eigenvalues 
$\lambda_s = +1$ (for $s=1,2$) or $\lambda_s = -1$ (for $s=3,4$) respectively. Note that $B$ is traceless. The normalization and the completeness condition are given by 
\begin{align}\label{wsnormalization}
 w_s^\dagger w_{s^\prime} = \delta_{ss^\prime},
 \qquad
 \sum_{s=1}^4 w_s w_s^\dagger = \bm 1. 
\end{align}
Charge conjugation operator $\mathcal C$ is defined by
\begin{equation}\label{defC}
\mathcal C\,{}^t\!\gamma^\mu \mathcal C^{-1} = -\gamma^\mu, 
\end{equation}
where ${}^t\!\gamma^\mu$ denotes the transpose of the gamma matrices, and the Hermitian/anti-Hermitian properties of 
the gamma matrices $(\gamma^\mu)^\dagger = \gamma^0\gamma^\mu\gamma^0$ indicate that 
$\mathcal C\, B^\ast \mathcal C^{-1} = \mathcal C\, {}^t\!B \mathcal C^{-1} = -B$. Then we find 
$B (\mathcal C w_s^\ast) = -\lambda_s (\mathcal C w_s^\ast)$. 
We specify 
\begin{equation}\label{chargeconjws}
w_1 = -\mathcal Cw_3^\ast,\, w_3 = \mathcal Cw_1^\ast,\, w_2 = -\mathcal Cw_4^\ast,\, w_4 = \mathcal Cw_2^\ast, 
\end{equation}
with 
$\mathcal C = -\mathcal C^{-1} = \mathcal C^\ast = -\mathcal C^\dagger.$

With the aid of the spatial homogeneity of the system we consider, 
we introduce the following decomposition of the solution for the Dirac equation: 
\begin{equation}
\zeta(\eta,\bm x) = \mathrm e^{-iHLz} \mathrm e^{i \bm k\cdot \bm x} \zeta_{\bm k,\,s}(\eta) w_s, 
\end{equation}
where we add a gauge fixing phase factor $\mathrm e^{-iHLz}$ for later convenience. 
The Schr\"odinger type equation for the mode function is then obtained as 
\begin{equation}\label{modeeqzeta}
\left( \partial_\eta^2 + \omega_k^2(\eta) - i\lambda_s\sigma(\eta) \,\right) \zeta_{\bm k,\,s}(\eta)=0, 
\end{equation}
where 
\begin{equation}
\omega_k^2(\eta) \equiv k^2 - 2aHLk_z + a^2H^2 (M^2+L^2), \quad
\sigma \equiv a^2H^2\sqrt{M^2+L^2}. 
\end{equation}
The two independent solutions for Eq.~(\ref{modeeqzeta}) is obtained in terms of the Whittaker functions 
$M_{\kappa,\mu^\pm}(z)$ and $W_{\kappa,\mu^\pm}(z)$. The parameters are given by 
\begin{equation}
\kappa = -iL\frac{k_z}{k},\;\; \mu^+ = \frac{1}{2}+i\sqrt{M^2+L^2},\;\; z = -2i\frac{k}{aH}, 
\end{equation}
for $s=1,2$ and 
\begin{equation}
\kappa = -iL\frac{k_z}{k},\;\; \mu^- = \frac{1}{2}-i\sqrt{M^2+L^2},\;\; z = -2i\frac{k}{aH}, 
\end{equation}
for $s=3,4$ respectively. 
To determine the positive frequency mode in the in-region ($\eta\to-\infty$), 
we can make use of an asymptotic formula of the Whittaker function \cite{gradshteyn2007}
$W_{\kappa,\mu^\pm}(z) \sim \mathrm e^{-z/2}z^\kappa$, and the positive frequency mode is given by 
\begin{equation}\label{zetain}
\zeta_{\bm k,\,s}^{+} = \frac{\mathrm e^{\pi i\kappa/2}}{\sqrt{2}k} 
\sqrt{\frac{1}{1-\frac{L}{\sqrt{L^2+M^2}}\frac{k_z}{k}}} W_{\kappa,\mu^\pm}(z)
\xrightarrow[z\to-i\infty]{\eta\to-\infty,\;a\to0} \frac{1}{\sqrt{2}k}\sqrt{\frac{1}{1-\frac{L}{\sqrt{L^2+M^2}}\frac{k_z}{k}}}\mathrm e^{-ik\eta}(-2k\eta)^{\kappa}, 
\end{equation}
where normalization is chosen to satisfy the canonical quantization condition as seen below. 
We can also find the negative frequency mode as 
\begin{equation}
\zeta_{\bm k,\,s}^{-} = \frac{\mathrm e^{\pi i\kappa/2}}{\sqrt{2}k}\sqrt{\frac{1}{1+\frac{L}{\sqrt{L^2+M^2}}\frac{k_z}{k}}} W_{-\kappa,\mu^\pm}(-z) 
\xrightarrow[z\to-i\infty]{\eta\to-\infty,\;a\to0} \frac{1}{\sqrt{2}k}\sqrt{\frac{1}{1+\frac{L}{\sqrt{L^2+M^2}}\frac{k_z}{k}}}\mathrm e^{+ik\eta}(-2k\eta)^{-\kappa}. 
\end{equation}
The transformation nature under the complex conjugation is given by 
\begin{equation}\label{zetaast}
(\zeta_{\bm k,\,s=1,2}^{\pm})^\ast = 
\sqrt{\frac{1\pm\frac{L}{\sqrt{L^2+M^2}}\frac{k_z}{k}}{1\mp\frac{L}{\sqrt{L^2+M^2}}\frac{k_z}{k}} } \zeta_{\bm k,\,s=3,4}^{\mp}. 
\end{equation}

From Eqs.~(\ref{chargeconjws}) and (\ref{zetaast}), we can obtain four independent solutions for 
the Dirac equation Eq.~(\ref{Dirac1st}) and construct the mode expansion for the quantized Dirac field $\hat\xi$ as 
\begin{equation}\label{modeexp}
\hat\xi(\eta,\bm x) = \mathrm e^{-iHLz} \int \frac{\mathrm d^3k}{(2\pi)^3} \sum_{s=1,\,2}
\left[ \hat b_{\bm k,\,s}u_{\bm k,\,s}(\eta)\mathrm e^{i\bm k\cdot\bm x}+\hat d_{\bm k,\,s}^\dagger v_{-\bm k,\,s}(\eta)\mathrm e^{-i\bm k\cdot\bm x} \right], 
\end{equation}
with
\begin{equation}\label{uvspinors}
 u_{\bm k,\,s} = \hat{\mathcal D} \zeta_{\bm k,\,s}^{+} w_s, \quad 
 v_{\bm k,\,s} = \mathcal C u_{\bm k,\,s}^\ast, \quad 
 \hat{\mathcal D} = \gamma^0\left(i\partial_\eta - k_i\gamma^0\gamma^i + aH\sqrt{M^2+L^2}B\right), 
\end{equation}
where we only need $s=1,2$ components. 
The anti-commutation relations 
\begin{equation}\label{anticomrel}
\{\hat b_{\bm k,\,s},\,\hat b_{\bm k^\prime,\,s^\prime}^\dagger\} = \{\hat d_{\bm k,\,s},\,\hat d_{\bm k^\prime,\,s^\prime}^\dagger\} = (2\pi)^3\delta^{(3)}(\bm k-\bm k^\prime)\delta_{s,\,s^\prime} 
,\;\; \mathrm{others} = 0, 
\end{equation}
are imposed as usual. 
The conjugate momentum of the canonical Dirac field $\hat\xi$ is given by $\hat\pi(\eta,\bm x) = \dfrac{\delta S}{\delta \hat{\xi}^\prime}=i\hat\xi^\dagger$. 
Therefore we obtain the conventional canonical quantization condition $\{\hat\xi(\eta,\,\bm x),\,\hat\pi(\eta,\,\bm y)\} = i \delta^{(3)}(\bm x-\bm y)$ (see also Appendix \ref{apdx:spinorformula}).

\section{Schwinger-induced current}\label{Sec:inducedcurrent}

\subsection{Vacuum expectation value of spinor current}
The in-vacuum state $\ket 0$ is defined as a state that satisfies the condition 
$b_{\bm k,\,s}\ket 0 = d_{\bm k,\,s}\ket 0 = 0$ for all $\bm k$ and $s=1,\,2$. 
Then, using the mode decomposition Eq.~(\ref{modeexp}) and the anti-commutation relation Eq.~(\ref{anticomrel}), 
the expectation value of the spinor current operator $J^3$ (along $z$-axis) is expressed as 
\begin{equation}
\braket{J^3} 
= -e\Braket{0|\hat{\bar\xi} \gamma^3 \hat\xi |0} 
= -e\int \frac{\mathrm d^3k}{(2\pi)^3}\sum_{s=1,\,2}v_{\bm k,\,s}^\dagger \gamma^0\gamma^3 v_{\bm k,\,s}. 
\end{equation}
The spinors $v_{\bm k,\,s}$ are defined in Eq.~(\ref{uvspinors}). 
Since the matrix $B$ can be regarded as $\bm 1$ on $w_s$ or $-\bm 1$ on $\mathcal C w_s^\ast$ for $s=1,\,2$, 
the vacuum expectation value of the induced current can be computed as 
\begin{equation}\label{braketJz}\begin{split}
&-e\Braket{0|\hat{\bar\xi} \gamma^3 \hat\xi |0} \\
&= \frac{-2eL}{\sqrt{L^2+M^2}} \int \frac{\mathrm d^3k}{(2\pi)^3}
 \left\{ {\zeta^+}^\prime{\zeta^+}^{\ast\prime} 
+i\left(\gamma k_z - F_k\right)({\zeta^+}{\zeta^+}^{\ast\prime} - {\zeta^+}^\prime{\zeta^+}^{\ast}) 
+ \left(2F_k^2 - \omega_k^2 - 2\gamma F_k k_z \right)|\zeta^+|^2 \right\} \\
&= \frac{-2eL}{\sqrt{L^2+M^2}} \int \frac{\mathrm d^3k}{(2\pi)^3}
\left\{ 1 + i\gamma k_z({\zeta^+}{\zeta^+}^{\ast\prime} - {\zeta^+}^\prime{\zeta^+}^{\ast}) 
+ 2 \left(F_k^2 - \omega_k^2 - \gamma F_k k_z \right)|\zeta^+|^2 \right \}, 
\end{split}\end{equation}
where 
\begin{equation}
\gamma \equiv \dfrac{\sqrt{L^2+M^2}}{L}-\dfrac{L}{\sqrt{L^2+M^2}},\quad 
F_k \equiv \dfrac{\omega_k \omega_k^\prime}{\sigma} = aH\sqrt{L^2+M^2}-\dfrac{Lk_z}{\sqrt{L^2+M^2}}, 
\end{equation}
and we have used the normalization condition Eq.~(\ref{normalizationzeta2}) (see Appendix \ref{apdx:spinorformula}) in the last line. 
Clearly, this integral 
diverges in the ultraviolet(UV) region ($k\to\infty$) and some renormalization procedure is required. In this paper, we apply the adiabatic subtraction method.

\subsection{Adiabatic subtraction}

The adiabatic subtraction is a renormalization scheme with which one subtracts the lower-order parts in the adiabatic 
(WKB) expansion of a quantity from its unrenormalized calculation result. 
The leading term in the adiabatic expansion of the expectation value of the current $-e\braket{0|\hat{\bar\xi} \gamma^3 \hat\xi |0}\bigl|^{(A)}$ 
imitates the divergence(s) in the UV (large $k$) region in the momentum space. 
Here, $-e\braket{0|\hat{\bar\xi} \gamma^3 \hat\xi |0}\bigl|^{(A)}$ is obtained by replacing the mode function $\zeta^+$ 
by the adiabatically (WKB) expanded counterpart $\zeta^+ \bigl\vert^{(A)}$. 
Subtracting this quantity from the formally divergent expectation value $-e\braket{0|\hat{\bar\xi} \gamma^3 \hat\xi |0}$, one obtains 
the renormalized expectation value of the current operator. 
We perform these calculations in this subsection (see also Appendix \ref{apdx:spinorWKB}).

In this subsection, we recover $\hbar$ to make things much clearer. 
The equation of motion for $\zeta = \zeta_{\bm k,\,s=1,2}^+$ is, again, given by 
\begin{equation}\label{eomzetawithhbar}
\left( \hbar^2 \partial_\eta^2 + \omega_k^2(\eta) - i\hbar\sigma(\eta) \,\right) \zeta(\eta)=0. 
\end{equation}
Because the $-i\sigma$ term in the equation above comes from first-order derivative, we have to assign $\hbar^1$ in front of it. 
This odd order term is peculiar to the spinor case (does not appear in the scalar case), and the usual WKB ansatz, 
which is valid for the scalar mode function, 
\begin{equation}
\zeta \stackrel{!}{=} \frac{1}{\sqrt{\Omega_k(\eta)}}\mathrm e^{- \frac{i}{\hbar} \int\mathrm d\eta^\prime \Omega_k(\eta^\prime)}, 
\end{equation}
is inappropriate ($\Omega_k$ is a function to be determined as a power series of $\hbar$). 
Instead, the WKB ansatz for spinor should take the following form 
\cite{PhysRevD.91.064031, PhysRevD.89.044030, PhysRevD.91.124075, PhysRevD.45.4659} (see Appendix B for the derivation), 
\begin{equation}\label{WKBansatz}
\zeta = \sqrt{\frac{\sigma}{2\omega^2(\sigma+\omega^\prime)}}(1+\hbar F^{(1)}+\hbar^2 F^{(2)}+\cdots)\, \mathrm e^{- i/\hbar \int\mathrm d\eta^\prime (\omega+\hbar\omega^{(1)}+\hbar^2\omega^{(2)}+\cdots)}, 
\end{equation}
where $F^{(i)}$s and $\omega^{(i)}$s are (real) unknown functions to be determined by the equation of motion Eq.~(\ref{eomzetawithhbar}) 
and the normalization condition 
\begin{equation}\label{zetanormalization}
\hbar^2 \zeta^\prime (\zeta^\ast)^\prime - i\hbar F_k(\zeta (\zeta^\ast)^\prime - \zeta^\prime \zeta^\ast) + \omega_k^2 |\zeta|^2 = 1, 
\end{equation}
where we have recovered $\hbar$. 
Note that this normalization condition can be satisfied only perturbatively (in an order-by-order manner). 
More detailed explanation on this ansatz is shown in Appendix \ref{apdx:spinorWKB}. 
We find all the odd order terms vanish, that is, $F^{(1)} = F^{(3)} = \cdots = 0$ and $\omega^{(1)} = \omega^{(3)} = \cdots = 0$. 
We can express $F^{(i)}$ and $\omega^{(i)}$ in terms of $\omega$ and $\sigma$. 
For example, at the second order they read 
\begin{equation}
\omega^{(2)} = -\frac{\sigma+\omega^\prime}{8 \sigma}\, \frac{\sigma^2+2\omega\sigma^\prime-5\omega^\prime\sigma}{\omega^3},\quad 
F^{(2)} = -\frac{\sigma+\omega^\prime}{16 \sigma}\,\frac{5\omega^\prime\sigma - 2\omega \sigma^\prime}{\omega^4}.
\end{equation}

With the adiabatic subtraction method, the renormalized current is given by 
\begin{equation} \label{J3 ren}
\Braket{0|J^3|0}_{\mathrm{ren}} 
= \Braket{0|J^3|0} - \Braket{0|J^3|0}\biggl|^{(2)},
\end{equation}
where $|^{(2)}$ means that the second term in the right hand side includes the contribution up to adiabatic order two. 
We calculate the momentum integral in the following way. 
First we introduce a momentum cutoff $\Lambda$ to control the divergences. 
Second, the momentum integrals of the exact part (the first term in Eq.~\eqref{J3 ren}) 
and the adiabatic part (the second term) are computed separately. Third, the subtraction is done, 
while the momentum cutoff $\Lambda$ is kept finite. Finally, we take the limit $\Lambda \to \infty$ and obtain a finite result.

The detailed calculation of the integrals can be found in appendix \ref{apdx:integration}, 
and here we show the final results. The first term in Eq.~\eqref{J3 ren} is given by 
\begin{equation}\label{formalJ3}\begin{split}
& \braket{J^3}= -2eL(aH)^3 \lim_{\Lambda\to\infty}\biggl [ \frac{1}{6\pi^2}\left(\frac{\Lambda}{aH}\right)^2 - \frac{1}{6\pi^2}\ln\left(\frac{2\Lambda}{aH}\right) \\ 
& + \frac{7}{72\pi^2}-\frac{L^2}{15\pi^2}-\frac{M^2}{12\pi^2} -\frac{3rM^2}{8\pi^2L^2} - \frac{3M^2 r}{16\pi^2L^3}\log\left(\frac{r-L}{r+L}\right) \\ 
& -\frac{r\mathrm{csch}(2\pi r)}{48\pi^5 L^2 }\left\{(45-\pi^2(11-12L^2+8r^2))\cosh(2\pi L) - (45-\pi^2(11-72L^2+8r^2))\frac{\sinh(2\pi L)}{2\pi L} \right\} \\ 
& + \frac{3rM^2 \mathrm{csch}(2\pi r)}{32\pi^2 L^3}\sum_{s=\pm} s\mathrm e^{2\pi rs}(\mathrm{Ei}(2\pi s(r+L)) - \mathrm{Ei}(2\pi s(r-L))) \\ 
& + \frac{\mathrm{csch}(2\pi r)}{16\pi^2} \Re[\int_{-1}^1\mathrm dx (1+r^2-(1+3L^2+3r^2)x^2+5L^2x^4)
\sum_{s=\pm}s(\mathrm e^{2\pi Lx} - \mathrm e^{-2\pi rs})\psi(i(Lx+rs)) ] \biggl ], 
\end{split}\end{equation}
with $r \equiv \sqrt{L^2+M^2}$. $\mathrm {Ei}(z)$ is the exponential integral function defined as 
$\mathrm {Ei}(z) = -\mathcal P\int_{-z}^{\infty}\mathrm dt \mathrm e^{-t}/t$ ($\mathcal P$ denotes Cauchy's principal value) 
and $\psi(z) = (\ln \Gamma(z))^\prime$ is the digamma function. 
We can see there are the quadratic and the logarithmic divergences in momentum cutoff $\Lambda$. 
The second term in Eq.~\eqref{J3 ren} (the subtraction term) is given by 
\begin{equation} \label{subtractJ3}
\Braket{0|J^3|0}\biggl|^{(2)} 
= -\lim_{\Lambda\to\infty}2eL \left\{
\left(\frac{aH\Lambda^2}{6\pi^2}-\frac{(aH)^3(2L^2+5M^2)}{60\pi^2}\right)\hbar^0 
+ \frac{(aH)^3}{18\pi^2}\left(4-3\ln\left(\frac{2\Lambda}{aHM}\right)\right)\hbar^2 \right\}, 
\end{equation}
where $\hbar$ is a constant which is taken to be a small expansion parameter in the adiabatic expansion and set to be unity after the truncation. 
It should be noted that the divergent parts of Eqs.~\eqref{formalJ3} and \eqref{subtractJ3} are exactly the same. 
Therefore, after the subtraction, we obtain the renormalized expectation value of the induced current as 
\begin{equation}\label{Jzren}\begin{split}
&\Braket{J^3}_{\mathrm{ren}} = \frac{eL(aH)^3}{4\pi^2}\biggl [
1 + \frac{4L^2}{15} + \frac{4}{3}\log M + \frac{3M^2}{L^2}\left(1 + \frac{r}{2L}\log\left(\frac{r-L}{r+L}\right)\right)  \\ 
& + \frac{r\mathrm{csch}(2\pi r)}{6\pi^3 L^2 }\left\{(45-\pi^2(11-12L^2+8r^2))\cosh(2\pi L) - (45-\pi^2(11-72L^2+8r^2))\frac{\sinh(2\pi L)}{2\pi L} \right\} \\ 
& - \frac{3rM^2 \mathrm{csch}(2\pi r)}{4L^3}\sum_{s=\pm} s\mathrm e^{2\pi rs}(\mathrm{Ei}(2\pi s(r+L)) - \mathrm{Ei}(2\pi s(r-L))) \\ 
& - \frac{\mathrm{csch}(2\pi r)}{2} \int_{-1}^1\mathrm dx (1+r^2-(1+3L^2+3r^2)x^2+5L^2x^4)
\sum_{s=\pm}s(\mathrm e^{2\pi Lx} - \mathrm e^{-2\pi rs})\Re\psi(i(Lx+rs)) \biggl ]. 
\end{split}\end{equation}

\section{Implications of the result}\label{Sec:result}
\begin{figure}[htb]
 \begin{center}
  \includegraphics[width=150mm]{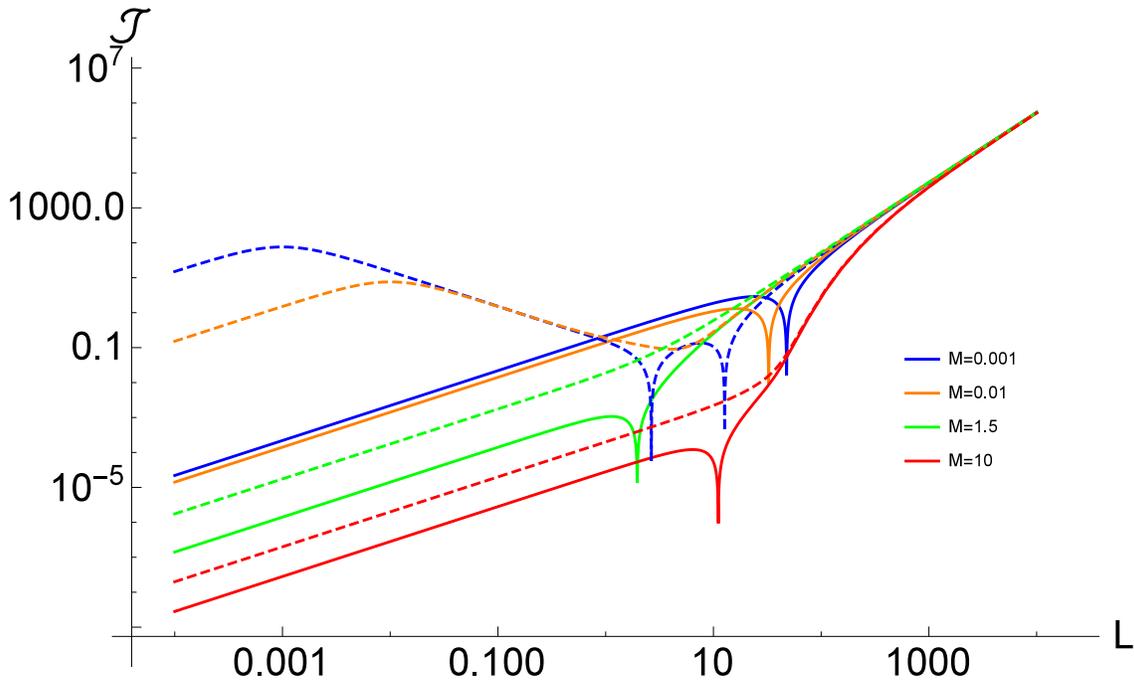}
 \end{center}
 \caption{The renormalized spinor current (solid) and the doubled scalar current (dashed) 
 induced by Schwinger effect in $1+3$ de Sitter space are shown. 
 The horizontal axis denotes the strength of the electric field $L\equiv eE/H^2$. 
 The mass parameter of the charged particles are $M\equiv m/H =10^{-3}$(blue), $10^{-2}$(orange), $1.5$(green) and $10$(red). 
 The absolute value $|\mathcal{J}|$ is plotted and 
 its sign flips around $L\sim 10$ in the spinor case with any mass and the scalar case with a sufficiently small mass. }
\label{fig:renormalized current}
\end{figure}

We investigate the renormalized current Eq.~(\ref{Jzren}) in this section. 
To this end, we introduce a dimensionless quantity $\mathcal{J}$ 
which is a function of the two dimensionless parameters 
$L=eE/H^2$ (electric field strength) and $M=m/H$ (spinor mass),
\begin{equation}
\mathcal{J}(L,M) \equiv \frac{|\braket{J^3}_{\mathrm{ren}}|}{ea^3H^3}. 
\end{equation}
In Fig.~\ref{fig:renormalized current}, behaviors of the spinor current $\mathcal{J}(L,M)$ (solid line) and 
the corresponding scalar current found in \cite{kobayashi2014schwinger} (dashed line) are shown. 

We have doubled the value of the bosonic current for comparison, since the spinor has two spin degrees of freedom. 
We can see the agreement between the bosonic and fermionic current in the strong electric field region $L\gg1, M^2$. 
However, they show different behaviors for $L\lesssim 1, M^2$.

\subsection{General property}
$\mathcal{J}(L,M)$ has some remarkable properties. 
The most intelligible one is an antisymmetry $\mathcal{J}(L,M) = -\mathcal{J}(-L,M)$ and its consequence $\mathcal{J}(0,M)=0$. 
This means that the renormalized current always vanishes at $L=0$ as expected. 
However, $\mathcal{J}(L,M)$ also becomes zero 
at a certain positive $L$ depending on $M$, namely $L_\ast(M)>0$, for any value of the mass parameter $M$. 
The spinor renormalized current is positive for $L>L_\ast$, and is negative for $L<L_\ast$. 
The negative spinor current is always (for any value of $M$) observed in the weak electric field regime 
in contrast with the bosonic case which shows the negative current only in the small mass regime $M_{\mathrm{scalar}}\lesssim0.003$. 

Another striking difference between the bosonic and the fermionic current is the absence of the IR hyperconductivity 
which is the rapid growth of $|\mathcal J|$ for smaller $L$. 
The hyperconductivity happens only in IR regime ($L<1$ and $M<1$).  
It was first found and discussed for the bosonic current in the $1+1$ dimensional de Sitter spacetime \cite{1475-7516-2014-04-009} 
and that in the $1+3$ dimensional de Sitter spacetime \cite{kobayashi2014schwinger}. 
In Fig.~\ref{fig:renormalized current}, one can see it as the peak of the bosonic current (dashed line) 
at $L\approx 10^{-3}$ and $10^{-2}$ for $M=10^{-3}$ and $10^{-2}$, respectively. 
Nevertheless, we find that there is not such a peak or the enhancement of the fermionic current in the IR regime 
except for a weak logarithmic divergence $\log m$ (see the next subsection for details). 
The absence of the hyperconductivity of the fermionic current in the $1+1$ dimensional de Sitter spacetime is also reported in \cite{PhysRevD.93.025004}.

\subsection{Strong and weak field limits}
In the strong electric field limit, $L\gg1,\,M$, the second line of Eq.~(\ref{Jzren}) dominates the expectation value and we obtain (for $L>0$) 
\begin{equation}\label{stronglimit}
\mathcal{J} \simeq \frac{L^2}{6\pi^3}\mathrm e^{-\frac{\pi M^2}{L}} = H^{-4}\frac{(eE)^2}{6\pi^3} \mathrm e^{-\frac{\pi m^2}{eE}}, 
\end{equation}
where the famous suppression factor of Schwinger effect in Minkowski spacetime $\exp(-\pi m^2/eE)$ is reproduced. 
We can also find the quadratic behavior ($\mathcal{J} \sim L^2$) of the renormalized current which is the same as the scalar (bosonic) current. 
This strong electric field limit corresponds to the Minkowski (weak curvature) limit $H\to0$. 
Thus, there is $H^{-1}$ divergence in $\braket{J}_{\mathrm {ren}}$ in this limit. 
This is caused by the lack of cosmic dilution in this limit. 
The particle produced at $t=-\infty$ contributes to the current forever. 
If we regulate the $H^{-1}$ divergence by the cosmic time interval $(t-t_0)$ with $t_0$ being the turn-on time of the electric field, 
we obtain $\braket{J}_{\mathrm {ren}} \sim e^3 E^2 (t-t_0)\exp(-\pi m^2/eE)$. 
This linear growth in time is consistent with the previous work of Schwinger effect in Minkowski spacetime shown in \cite{Anderson:2013ila}. 

In contrast to the intuitive behavior in the strong field limit, 
the strange negativity of the renormalized current appears in the weak electric field regime $L\ll1$. 
In this limit, Eq.~(\ref{Jzren}) becomes 
\begin{equation}
\mathcal{J} \simeq \frac{L}{3\pi^2} \left[ \log M - \Re \psi(iM) - \frac{\pi M(4M^2+1)}{3\sinh(2\pi M)} \right]. 
\end{equation}
We can define a dimensionless conductivity in the weak electric field limit as $\sigma(M) = \mathcal J(L,M)/L|_{L\to0}$ which is negative for all $M$. 
The massless limit of the conductivity is given by 
\begin{equation}
\sigma(M) \xrightarrow{M\to0} \frac{1}{3\pi^2} \log M + \frac{6\gamma_E-1}{18\pi^2} + \mathcal O(M^2), 
\end{equation}
where $\gamma_E$ is the Euler constant. 
There is no power-law IR enhancement but the logarithmic divergence for the spinor conductivity. 
On the other hand, the bosonic conductivity has a much faster enhancement in proportional to $M^{-2}$, which is called the IR hyperconductivity, in the small mass limit, $M\to 0$. 
The massive limit is given by 
\begin{equation}\label{sigmamassive}
\sigma(M) \xrightarrow{M\to\infty} \left(-\frac{1}{36\pi^2 M^2} + \mathcal O(M^{-4}) \right) - \frac{2}{9\pi}\mathrm e^{-2 \pi M}M(4M^2+1). 
\end{equation}
As expected, $\sigma(M)$ is suppressed in the massive limit. 
Schwinger mechanism cannot produce massive fermions effectively due to the suppression factor $\exp(-\pi m^2/eE)$. 
The gravitational particle production is also suppressed by the factor $(\exp(2\pi M)+1)^{-1} \sim \exp(-2\pi M)$. 
Thus we might be able to identify the latter term in Eq.~(\ref{sigmamassive}) as the effect of the gravitational particle production. 
Nevertheless, we do not have any satisfactory explanation for its negative sign. 
Moreover, the terms in the parenthesis in Eq.~(\ref{sigmamassive}) do not have exponential suppression factors 
and cannot be simply attributed to either of Schwinger or the gravitational particle production. 
Their origins are still unidentified but we further discuss it in the next subsection. 
Here, we note that the higher order adiabatic subtraction can remove 
some of these terms with the power-law dependence on $M$. 
For instance, the first term with $M^{-2}$ in Eq.~(\ref{sigmamassive}) can be removed by the adiabatic subtraction of order $\hbar^4$, 
while it adds a new $\mathcal O(L^3/M^2)$ term to the induced current and changes the IR behavior of the current. 
Nevertheless, the higher order ($\mathcal O(M^{-4})$) terms which are not suppressed by the exponential factor in \eqref{sigmamassive} still remain even in this case.

\subsection{Negativity of the induced current}
It is questionable whether we should take the strange negativity of the current seriously. 
The range of wavelength which is short enough to verify the adiabatic subtraction depends on the particle mass, 
and only the modes with $(k/aH)^2+M^2 \gg 1$ can imitate the correct behavior of the exact mode function. 
Thus, the adiabatic approximation is not necessarily correct for the long wavelength modes when $m \ll H$.
A possible criticism is that the adiabatic expansion is inappropriate 
for fields with extremely small masses and the adiabatic subtraction scheme becomes 
invalid for the modes with $m/H \ll k/(aH) \ll 1$ though they are in the UV regime. 

However, it has been confirmed in \cite{THJY2016} that the point splitting renormalization scheme is in perfect accord with 
the adiabatic subtraction for the scalar current. 
This implies that the strange behaviors we have found in the previous section have nothing to do with the accuracy of the WKB expansion in infrared regime. 
Thus, it is worthwhile to investigate physical consequences of the result, Eq.~\eqref{Jzren}, in this section. 

The semiclassical equation of motion for the gauge field is given by $F^{\mu\nu}_{\;\;\; ,\nu} = \braket{J^\mu}_{\mathrm{ren}}$ 
in our convention. For the electric background field $E_z = -A_z^{\,\prime}$, the equation of motion is given by 
\begin{equation}
E_z^{\,\prime} = -\braket{J^3}_{\mathrm{ren}}, 
\end{equation}
which can be regarded as a feedback system. 
It is easy to figure out the stability of the electric field-current system by looking at the signature of the renormalized current. 
The positive current reduces the background electric field while the negative current enhances it. 
The zeros of the current correspond to either a stable point or a saddle (unstable) point. 

\begin{figure}[htb]
 \begin{center}
  \includegraphics[width=110mm]{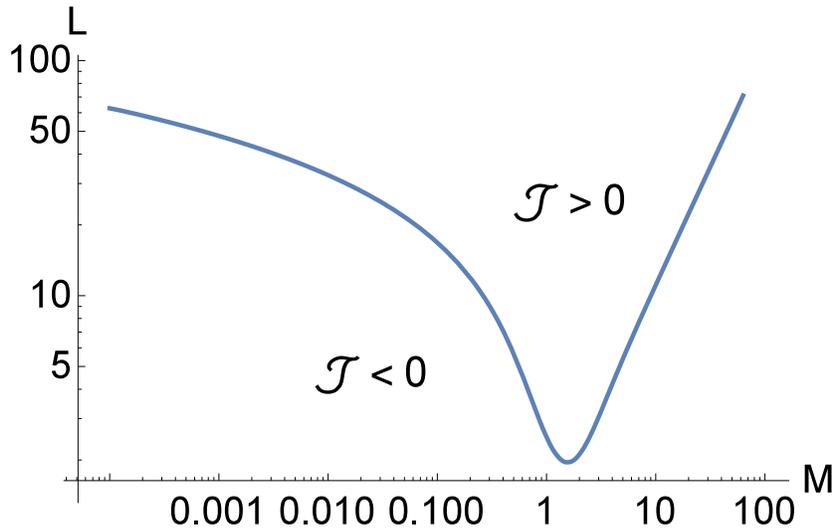}
 \end{center}
 \caption{Zero of the renormalized current $\mathcal{J}(L,M)$ in $L$-$M$ plain. 
 The upper region corresponds to the positive current $\mathcal{J}>0$ (negative feedback) and 
 the lower region corresponds to the negative current $\mathcal{J}<0$ (positive feedback). 
 The line shows positions of the stable points of the electric field-current system. }
 \label{fig:zeroofJ}
\end{figure}

Surprisingly enough, the trivial zero $L=0$ ($E=0$) is not a stable point but a saddle point. 
This situation is opposite of the case of the scalar current. 
Another zero $L = L_\ast >0$ is always a stable point. We plot $L_\ast$ as a function of $M=m/H$ in Fig.~\ref{fig:zeroofJ}. 
This figure can be seen as a phase diagram of the system. 
The negative current occurs in the region below the blue line. 
A similar diagram for the scalar current is discussed also in \cite{THJY2016}. 
Note that the negativity of the induced current is not past redemption even though it indicates the instability of the system. 
This is not a bottomless instability since the current becomes positive for a sufficiently strong electric field.


\section{Conclusion}\label{Sec:conclusion}

In this paper, we have investigated the fermionic current induced by Schwinger effect in $1+3$ dimensional de Sitter spacetime. 
We have considered a homogeneous electric  field Eq.~(\ref{backgroundAz}) which has the constant energy density in the expanding universe. Using the adiabatic subtraction, we obtained the renormalized expectation value of the current operator for the charged fermion. 
The analytic result Eq.~\eqref{Jzren} was studied in detail and the similarity and difference from the bosonic (scalar) case were discussed. 

With the aid of the analytic result Eq.~\eqref{Jzren}, we managed to investigate the 
behaviors of the induced current 
in both the weak and strong electric field limits. In the strong field limit, we obtained Eq.~\eqref{stronglimit} which coincides with 
the behavior of the bosonic current \cite{kobayashi2014schwinger} as well as that of the current in flat spacetime \cite{Anderson:2013ila}. 
Since the contribution to the induced current mainly comes from Schwinger pair production in strong field regime, 
the induced current in this regime 
carries the mass suppression factor for the Schwinger effect $\exp(-\pi m^2/eE)$ as expected. 
On the other hand, in the weak electric field regime, we have found two remarkable features, namely
the absence of the IR hyperconductivity 
and the negativity of the induced current. 
These features have been observed in the $1+1$ dimensional fermionic case \cite{PhysRevD.93.025004}
and the $1+3$ dimensional bosonic case \cite{kobayashi2014schwinger}, respectively. 
In our case, the negative current occurs for the electric field smaller than a certain value $L_\ast(m/H)$ determined by the spinor mass as plotted in Fig.~\ref{fig:zeroofJ}. 
Although the negativity of the current indicates the positive feedback which counter-intuitively enhances the background electric field, 
it does not mean an unbounded 
instability. 
The system is stable for electric field which is stronger than $L_*(m/H)$ and thus the electric field is not enhanced 
beyond $L_*$ due to the instability. 
We also found the terms which do not carry any exponential mass suppression factor 
in the massive spinor limit of the conductivity Eq.~\eqref{sigmamassive}. 
If the particle is sufficiently heavy, 
the semiclassical description must be precise and it suggests that the exponential mass suppression factors such as 
$\exp(-\pi m^2/eE)$ or $\exp(-2\pi m/H)$ should appear. 
Thus, these terms indicate contributions beyond the semiclassical approximations. 
Further investigation is needed to clarify the origin or the physical 
interpretation of the negative current and the terms without the exponential mass suppression. 
It should be noted that the expression for the renormalized current Eq.~\eqref{Jzren} apparently has 
a logarithmic divergence in the massless limit 
which has been introduced by the adiabatic subtraction,
while the current vanishes in the massive limit without divergence. 

Our result is a significant step towards understanding the electromagnetic response of the inflationary spacetime. 
However, we need further study to obtain  general implications from Schwinger effect for inflationary magnetogenesis. 
This is because in this paper we have restricted ourselves to the following two points: 
(i) We have focused only on the specific background gauge field configuration which scales as $A_z \propto (a-1),$
though the kinetic coupling model produces a gauge field with $A_i \propto a^s$ ($s$ is a real parameter of the model)~\cite{martin2008generation} 
which may not be approximated by the setup in the present paper. 
(ii) We have not considered the realistic dynamics of the gauge field with the backreaction effect from the particle production. 

\acknowledgments
This work was supported by JSPS KAKENHI, 
Grant-in-Aid for JSPS Fellows 15J09390 (TH), 
Grant-in-Aid for Scientific Research 15H02082 (JY), 
Grant-in-Aid for Scientific Research on Innovative Areas 15H05888 (JY). 
TH would like to thank Teruaki Suyama for his instructive notebook on the spinor in curved spacetime. 
The work of TF has been supported in part by the JSPS Postdoctoral Fellowships for Research Abroad (Grant No. 27-154).

\appendix 
\section{Some formulae for spinor calculation}\label{apdx:spinorformula}
We need a spin sum formula to calculate the anti-commutation relation $\{\hat\xi(\eta,\,\bm x),\,\hat\pi(\eta,\,\bm y)\}$. 
What should be computed is a quantity $\displaystyle\sum_{s=1,\,2} (u_{\bm k,\,s} u_{\bm k,\,s}^\dagger+v_{\bm k,\,s} v_{\bm k,\,s}^\dagger)$. 

Let $X$ be a matrix constructed from the eigenvectors $w_s$ in Eq.~(\ref{defws}) as 
\begin{equation}
X \equiv \sum_{s=1,\,2} w_sw_s^\dagger. 
\end{equation}
$X$ should be a Hermitian matrix ($X^\dagger = X$) due to the orthogonality of the eigenvectors $w_s$. 
The completeness condition reads $X + \mathcal C X^\ast \mathcal C^\dagger = \bm 1$. $X$ also satisfies a condition $BX=X$, 
where $B$ is defined  as $B\equiv \dfrac{1}{\sqrt{M^2+L^2}}(M\gamma^0+L\gamma^0\gamma^3)$. 
One can show that the unique representation for $X$ in terms of gamma matrices is 
\begin{equation}
X = \frac{1}{2} + \frac{M}{2\sqrt{M^2+L^2}}\gamma^0 + \frac{L}{2\sqrt{M^2+L^2}}\gamma^0\gamma^3 = \frac{1}{2}(1+B). 
\end{equation}
Equation \eqref{defC} leads to $\mathcal C X^\ast \mathcal C^\dagger = \frac{1}{2}(1-B)$, and the completeness condition is manifestly satisfied. 
After some algebra, one can find 
\begin{equation}\label{normalizationzeta}
\sum_{s=1,\,2} (u_{\bm k,\,s} u_{\bm k,\,s}^\dagger+v_{\bm k,\,s} v_{\bm k,\,s}^\dagger )
= \left[ \zeta^\prime (\zeta^\ast)^\prime 
- iF_k(\zeta (\zeta^\ast)^\prime - \zeta^\prime \zeta^\ast) + \omega_k^2 |\zeta|^2 \right] \bm 1, 
\end{equation}
where $\zeta$ is a shorthand notation for $\zeta_{\bm k,\,s=1,2}^{+}$, the suffix $s$ is indeed verbose since $\zeta_{\bm k,\,s=1}^{+}=\zeta_{\bm k,\,s=2}^{+}$, 
and $F_k \equiv \dfrac{\omega_k \omega_k^\prime}{\sigma}$. 
Using the equation of motion Eq.~(\ref{modeeqzeta}) and the normalization of Eq.~(\ref{zetain}), 
we find that the large parenthesis in Eq.~(\ref{normalizationzeta}) equals to unity. Therefore we obtain the spin sum formula as 
\begin{equation}
\sum_{s=1,\,2} (u_{\bm k,\,s} u_{\bm k,\,s}^\dagger+v_{\bm k,\,s} v_{\bm k,\,s}^\dagger ) = \bm 1. 
\end{equation}

We can express this normalization condition for the mode function $\zeta$ in a simpler way by introducing 
an auxiliary 
function $\tilde \zeta \equiv (\omega_k^2 - F_k^2)^{-1/2}(\partial_\eta - iF_k)\zeta$ as 
\begin{equation}\label{normalizationzeta2}
|\zeta|^2 + |\tilde \zeta|^2 = \frac{1}{\omega_k^2 - F_k^2},
\end{equation}
where $\omega_k^2 - F_k^2 = k^2 - \dfrac{L^2}{L^2+M^2}k_z^2$ is a time independent constant. 
Note also that $F_k^\prime = \sigma$ and $\zeta = -(\omega_k^2 - F_k^2)^{-1/2}(\partial_\eta + iF_k)\tilde\zeta$. 

In our convention, the normalization condition for $u,\,v$ spinors is expressed as 
\begin{equation}
u_{\bm k,\,s}^\dagger u_{\bm k,\,s^\prime} = v_{\bm k,\,s}^\dagger v_{\bm k,\,s^\prime} = \delta_{s,\,s^\prime}, 
\end{equation}
and we can also check the orthogonality condition for $s,\,s^\prime=1,\,2$ 
\begin{equation}
u_{\bm k,\,s}^\dagger v_{\bm k,\,s^\prime} = v_{\bm k,\,s}^\dagger u_{\bm k,\,s^\prime} = 0. 
\end{equation}

Let us also write down a useful formula which is needed in calculation of the expectation value of 
the current operator (\ref{braketJz}), for $s=1,\,2$, 
\begin{equation}
w_s^\dagger \gamma^3\gamma^0 w_s = w_s^\dagger \frac{B\gamma^3\gamma^0 + \gamma^3\gamma^0 B}{2} w_s = -\frac{L}{\sqrt{L^2+M^2}},
\end{equation}
where we used $Bw_s=w_s$ for $s=1,\,2$ and $B^\dagger = B$.

\section{Adiabatic expansion for spinor mode function}\label{apdx:spinorWKB}
In order to find the consistent WKB expansion Eq.~(\ref{WKBansatz}) for the equation of motion Eq.~(\ref{eomzetawithhbar}), 
we examine the most primitive WKB-type ansatz such as $\zeta = \exp({\pm\hbar^{-1}\int^\eta\mathrm d\eta^\prime(X(\eta^\prime)+iY(\eta^\prime))})$ 
where $X$ and $Y$ are real functions to be determined by the equation of motion. 
Substituting it into Eq.~(\ref{eomzetawithhbar}), one finds 
\begin{equation}\label{eomXY}
X^2-Y^2\pm \hbar X^\prime = -\omega^2,\quad 2XY\pm \hbar Y^\prime = \hbar \sigma.
\end{equation}
Note that 
$X$ ($Y$) has only odd (even) order terms in the power series expansion of $\hbar$, respectively. 
The latter of Eq.~(\ref{eomXY}) reads $X = \mp \frac{1}{2}(\ln Y)^\prime + \frac{\sigma}{2Y}$. 
We can eliminate $X$ and introduce a normalization factor $\mathcal N$ to rewrite the ansatz as 
$\zeta = \dfrac{\mathcal N}{\sqrt{Y}}\mathrm e^{{\pm \frac{i}{\hbar}\int^\eta\mathrm d\eta^\prime(Y - i\hbar\frac{\sigma}{2Y})}}$. 
Unfortunately, this expression does not satisfy the normalization condition Eq.~(\ref{zetanormalization}) in a nonperturbative way. 
However, the normalization condition can be satisfied at each order of the $\hbar$ expansion. 

At the zeroth order, the solution is found to be $Y=\omega$. 
The relation $F_k^\prime = (\frac{\omega\omega^\prime}{\sigma})^\prime = \sigma$ 
can be used to find an integral $\frac{\sigma}{\omega} = (\ln (\omega+F_k))^\prime$ and then we find that the normalized 
positive frequency mode is given by
\begin{equation}
\zeta^+|^{(0)} = \sqrt{\dfrac{\sigma}{2\omega^2(\sigma+\omega^\prime)}}\mathrm e^{-\frac{i}{\hbar}\int\mathrm d\eta^\prime \omega}.
\end{equation}
If we write $Y$ in a power series of $\hbar$ as 
$Y = \displaystyle \sum_{n=0}^{\infty} \hbar^n \omega^{(n)}$ with $\omega^{(0)} = \omega (= \omega_k(\eta))$ 
and also rewrite $\dfrac{\sigma}{2Y}$ term in the exponential as 
$\displaystyle \sqrt{\dfrac{\sigma}{2\omega^2(\sigma+\omega^\prime)}}\sum_{n=0}^{\infty} \hbar^n F^{(n)}$ with $F^{(0)}=1$ , 
we finally obtain the consistent expansion Eq.~(\ref{WKBansatz}). 
This expression gives correct asymptotic feature of the exact solution (the Whittaker function) as described in Eq.~(\ref{zetain}). 
On the other hand, we can also find that the zeroth order ansatz for the negative frequency mode function $\zeta^-|^{(0)}$ is given by 
$\zeta^-|^{(0)} = \sqrt{\dfrac{\sigma}{2\omega^2(\sigma-\omega^\prime)}} \mathrm e^{+\frac{i}{\hbar}\int\mathrm d\eta^\prime \omega}$, 
which is slightly different from the positive counterpart.

\section{Integration of the Whittaker function}\label{apdx:integration}
Here we describe the procedure to evaluate the integral Eq.~(\ref{braketJz}),
\begin{equation}\label{exactpartofcurrent}
\braket{J^3} = 
\frac{-2eL}{\sqrt{L^2+M^2}}\int \frac{\mathrm d^3k}{(2\pi)^3}
\Big\{1 + i\gamma k_z({\zeta^+}{\zeta^+}^{\ast\prime} - {\zeta^+}^\prime{\zeta^+}^{\ast}) - 2 \left(\omega_k^2-F_k^2+\gamma F_k  k_z \right)|\zeta^+|^2  \Big\}, 
\end{equation}
with a cutoff in momentum $\displaystyle \int_0^\infty\mathrm dk \to \lim_{\Lambda\to\infty}\int_0^\Lambda\mathrm dk$. 
The procedure is essentially the same as the previous works \cite{1475-7516-2014-04-009,kobayashi2014schwinger,PhysRevD.93.025004}. 
In the cylindrical coordinates, the above equation reads 
\begin{equation}\begin{split}
\frac{-2eL}{\sqrt{L^2+M^2}} &\lim_{\Lambda\to\infty} \int_0^\Lambda \frac{\mathrm dk}{2\pi}k^2 \int_{-1}^{1} \frac{\mathrm dx}{2\pi} \int_{0}^{2\pi} \frac{\mathrm d\varphi}{2\pi}  \\
&\Big\{ 1+i\gamma kx({\zeta^+}{\zeta^+}^{\ast\prime} - {\zeta^+}^\prime{\zeta^+}^{\ast}) - 2\left((1-x^2)k^2 + aH\sqrt{L^2+M^2}\gamma kx \right)|\zeta^+|^2  \Big\}, \end{split}\end{equation}
and the first trivial term gives a divergent contribution 
\begin{equation}\label{integtrivialpart}
\frac{-2eL}{\sqrt{L^2+M^2}} \lim_{\Lambda\to\infty} \int_0^\Lambda \frac{\mathrm dk}{2\pi}\int_{-1}^{1} \frac{\mathrm dx}{2\pi} \int_{0}^{2\pi} \frac{\mathrm d\varphi}{2\pi}
k^2 = \lim_{\Lambda\to\infty}\frac{-2eL\Lambda^3}{6 \pi^2 \sqrt{L^2+M^2}}. \end{equation}
The positive frequency mode function $\zeta^+$ defined in Eq.~(\ref{zetain}) is, again, given by 
\begin{equation}
\zeta_{\bm k}^+(\eta) = \frac{\mathrm e^{\frac{\pi}{2} Lx}}{\sqrt{2}k} \sqrt{\frac{1}{1-\frac{L}{\sqrt{L^2+M^2}}x}} W_{-iLx,\,\frac{1}{2}+i\sqrt{L^2+M^2}}\left(-2i\frac{k}{aH}\right). 
\end{equation}
For the remaining part, we can use the Mellin-Barnes type integral representation for the Whittaker function $W_{\kappa,\mu}(z)$ as is done in the previous works,
\begin{equation}
W_{\kappa,\mu}(z) =\int_{C_s}\frac{\mathrm ds}{2\pi i} z^s \mathrm e^{-z/2} \frac{\Gamma(s-\kappa)\Gamma(-s-\mu+\frac{1}{2})\Gamma(-s+\mu+\frac{1}{2})}{\Gamma(\frac{1}{2}-\kappa-\mu)\Gamma(\frac{1}{2}-\kappa+\mu)}, 
\end{equation}
where the contour $C_s$ runs from $-i\infty$ to $i\infty$ and is taken to separate the poles of $\Gamma(s-\kappa)$ ($s=\kappa-n$, $n=0,1,2,\cdots$) from 
the ones of $\Gamma(-s-\kappa-\mu+\frac{1}{2})\Gamma(-s-\kappa+\mu+\frac{1}{2})$. 
Using the complex conjugation nature $(W_{\kappa,\mu}(z))^\ast = W_{\kappa^\ast,\mu^\ast}(z^\ast)$, 
the differential property $\frac{\mathrm d}{\mathrm dz}W_{\kappa,\lambda}(z) = \left(\frac{1}{2}-\frac{\kappa}{z}\right)W_{\kappa,\lambda}(z) - \frac{1}{z}W_{1+\kappa,\lambda}(z)$
and the reflection formula for the Gamma function, the integral is rewritten as 
\begin{equation}\begin{split}
&-2eL\lim_{\Lambda\to\infty} 
\int_0^\Lambda \mathrm dk \int_{-1}^{1}\mathrm dx 
\int_{C_s}\frac{\mathrm ds}{2\pi i} \int_{C_t}\frac{\mathrm dt}{2\pi i} \frac{\mathrm e^{\pi Lx}}{4\pi^2}\,
\mathrm e^{\frac{\pi i}{2}(t-s)}\Gamma(s+iLx)\Gamma(-s-ir)\Gamma(-s+ir+1) \\
&\times \Gamma(t-iLx)\Gamma(-t+ir)\Gamma(-t-ir+1)\frac{\sinh{\pi(r-Lx)}\sinh{\pi(r+Lx)}}{\pi^2(r+Lx)}\left(\frac{2k}{aH}\right)^{s+t} \\
&\times \left\{ (x^2+\gamma x-1)k^2 - aH\gamma(r+Lx)(1+\frac{1}{2}\left(\frac{1+i(r-Lx)}{s+iLx-1}+\frac{1-i(r-Lx)}{t-iLx-1} \right))kx \right\}, 
\end{split}\end{equation}
where $r=\sqrt{L^2+M^2}$ and $\gamma = r/L-L/r$. 
The integration contours $C_s$ and $C_t$ run from $-i\infty$ to $+i\infty$. 
$C_s$ sees the poles at $s=-iLx-n$ ($n=0,1,2,\cdots$) on the left and the ones at $s=-ir+n,\,ir+1+n$ on the right. 
$C_t$ sees the poles at $t=+iLx-n$ on the left and the ones at $s=ir+n,\,-ir+1+n$ on the right. 

Since $C_s$ and $C_t$ can be taken to ensure $\Re(s+t)>0$, we can perform the $k$-integral explicitly 
to find $\mathcal O(\Lambda^{(s+t+2)})$ and $\mathcal O(\Lambda^{(s+t+3)})$ terms. 
We then perform the $t$-integral with closing the integration path positively (by a counterclockwise path). 
The residues from $t=iLx-m,\,(m=0,1,2,\cdots)$ and $t=-s-2,-s-3$ can contribute to the integral. 
Note that the contributions from $m>3$ poles will vanish after taking the limit $\Lambda\to\infty$. 
The $s$-integral can be similarly done for the contributions from $m=0,-1,-2,-3$, and remains only the residues from 
the poles at $s=-iLx,-iLx-1,-iLx-2,-iLx-3$. 
The non-vanishing contribution is calculated as 
\begin{equation}\label{divpart}\begin{split}
-2eL &\Bigl(-\frac{\Lambda^3}{6\pi^2r } + \frac{aH}{6\pi^2}\Lambda^2 + 0\times (aH)^2\Lambda^1 
- \frac{(aH)^3}{6\pi^2}\ln\left(\frac{2\Lambda}{aH}\right) + (aH)^3\mathcal O(\Lambda^0)\Bigl). 
\end{split}\end{equation}
We find the cancelation of the $\Lambda^3$ divergence here and in Eq.~(\ref{integtrivialpart}). The finite part ($\mathcal O(\Lambda^0)$ in Eq.~\eqref{divpart} 
) is given by 
\begin{equation}\label{Lambda0thcontribtion1}\begin{split}
\mathcal O(\Lambda^0) &= \frac{\gamma_E}{6\pi^2}-\frac{23}{144\pi^2}-\frac{7L^2}{120\pi^2}+\frac{L^4}{420\pi^2}-\frac{23M^2}{192\pi^2} 
-\frac{3M^2}{4\pi^2L^2}+\frac{L^2M^2}{1440\pi^2}-\frac{5M^4}{576\pi^2} \\
& -\frac{3M^2 r}{8\pi^2L^3}\log\left(\frac{r-L}{r+L}\right)
+i\left(\frac{1}{12\pi}-\frac{1}{2\pi^2 r}+\frac{121L^2}{216\pi^2r}-\frac{91L^4}{1440\pi^2r}+\frac{L^6}{2016\pi^2r}-\frac{65r}{144\pi^2} \right. \\ 
&\qquad\qquad\qquad\qquad\qquad\qquad
+\left. \frac{89rL^2}{1440\pi^2}-\frac{rL^4}{1120\pi^2}-\frac{41rM^2}{576\pi^2}-\frac{rM^2L^2}{1440\pi^2}-\frac{rM^4}{576\pi^2} \right)\\ 
&+\frac{1}{8\pi^2} \int_{-1}^1 \mathrm dx(1+r^2-(1+3L^2+3r^2)x^2+5L^2x^4)(\psi(iLx-ir)+\psi(iLx+ir)), 
\end{split}\end{equation}
where $\psi(z) = (\ln\Gamma(z))^\prime$ denotes the digamma function and $\gamma_E$ is the Euler constant. 
The $x$-integral cannot be expressed in terms of simpler functions, but it is real since the imaginary part of the digamma function 
is given by $2 \Im \psi(iy) = 1/y+\pi\coth(\pi y)$.

The other part (residues from $t=-s-2,-s-3$) is calculated as 
\begin{equation}\label{calcintegralfs}
-2eL(aH)^3\int_{-1}^1\mathrm dx\int_{C_s}\frac{\mathrm ds}{2\pi i}
\frac{\mathrm e^{-i\pi s}\mathrm e^{\pi Lx}\sinh(\pi(r-Lx))\sinh(\pi(r+Lx))}{\sin(\pi(s+iLx))\sin(\pi(s-ir))\sin(\pi(s+ir))}f(s),
\end{equation}
where $f(s)$ is meromorphic, and has only single poles located at $s = -iLx+1,-iLx,-iLx-1,-iLx-2,-iLx-3$. 
We further introduce a function 
\begin{equation}
g(s)=b_3(s+iLx)^3 + b_2(s+iLx)^2 + b_1(s+iLx) + \frac{c_0}{s+iLx} + \frac{c_1}{s+iLx+1} + \frac{c_2}{s+iLx+2} + \frac{c_3}{s+iLx+3}
\end{equation}
to express $f(s)$ as $f(s) = g(s) - g(s+1) + \dfrac{d}{s+iLx-1}$. 
All the coefficients $b_i,c_i,d$ have no $s$-dependence. 
The shift of the contour $s\to s-1$ does not change the coefficient in front of $f(s)$ in Eq.~(\ref{calcintegralfs}), 
then we find that the $(g(s)-g(s-1))$ part of (\ref{calcintegralfs})
\begin{equation}
\int_{C_s} \frac{\mathrm ds}{2\pi i} \cdots (g(s)-g(s-1)) 
= \left(\int_{C_s}-\int_{C_{s-1}}\right)\frac{\mathrm ds}{2\pi i} \cdots g(s), 
\end{equation}
is given by sum of the residues of the poles between $C_s$ and $C_{s-1}$, say, $s = -ir-1$, $s = ir$ and $s = -iLx+1$. 
The contributing poles of the $d$-term of (\ref{calcintegralfs}), 
\begin{equation}\label{dterminteg}
\int_{C_s}\frac{\mathrm ds}{2\pi i}
\frac{\mathrm e^{-i\pi s}\mathrm e^{\pi Lx} 
\sinh(\pi(r-Lx))\sinh(\pi(r+Lx))}{\sin(\pi(s+iLx))\sin(\pi(s-ir))\sin(\pi(s+ir))}\frac{d}{s+iLx-1}, 
\end{equation}
are $s = -iLx+2+n$, $s=ir+n+1$ and $s=-ir+n$ ($n = 0,1,2,\cdots$). 
These poles are negatively encircled by the integration path ($C_s+$(semicircle on the right half plane)), 
and the residue theorem gives 
\begin{equation}
-2eL(aH)^3\int_{-1}^1\mathrm dx \frac{d}{\pi} \sum_{n=0}^{\infty}
{\textstyle\left\{ -\frac{1}{1+n} 
+ \frac{\mathrm e^{\pi(r+Lx)}}{n+i(r+Lx)}\frac{\sinh(\pi(r-Lx))}{\sinh(2\pi r)} 
+ \frac{\mathrm e^{-\pi(r-Lx)}}{n-1-i(r-Lx)}\frac{\sinh(\pi(r+Lx))}{\sinh(2\pi r)}\right\} }. 
\end{equation}
Each sum $\displaystyle \sum_{n=0}^\infty \frac{1}{n+\alpha}$ ($\alpha \neq 0$) seems divergent, however, 
they are indeed finite, because using a series formula for the digamma function one can show 
\begin{equation}\begin{split}
& \sum_{n=0}^{\infty}{
\left\{ -\frac{1}{1+n} 
+ \frac{\mathrm e^{\pi(r+Lx)}}{n+i(r+Lx)}\frac{\sinh(\pi(r-Lx))}{\sinh(2\pi r)} 
+ \frac{\mathrm e^{-\pi(r-Lx)}}{n-1-i(r-Lx)}\frac{\sinh(\pi(r+Lx))}{\sinh(2\pi r)}\right\} } \\
& = \sum_{n=0}^{\infty}
\bigg\{  
\mathrm e^{\pi(r+Lx)}\frac{\sinh(\pi(r-Lx))}{\sinh(2\pi r)} \left(\frac{1}{n+i(r+Lx)}-\frac{1}{n+1}\right) \\
&\qquad\qquad\qquad\qquad\qquad\qquad +\mathrm e^{-\pi(r-Lx)}\frac{\sinh(\pi(r+Lx))}{\sinh(2\pi r)} \left(\frac{1}{n-1-i(r-Lx)}-\frac{1}{n+1}\right) \bigg\} \\
&= \mathrm e^{\pi(r+Lx)}\frac{\sinh(\pi(r-Lx))}{\sinh(2\pi r)} ( -\gamma_E-\psi(ir+iLx) ) \\ 
&\qquad\qquad\qquad\qquad\qquad\qquad +\mathrm e^{-\pi(r-Lx)}\frac{\sinh(\pi(r+Lx))}{\sinh(2\pi r)} (-\gamma_E-\psi(-1-ir+iLx) ). 
\end{split}\end{equation}
This result is also achieved by just applying the (Hurwitz's type) $\zeta$-function regularization technique to each of the sum. 

The contribution of Eq.~(\ref{calcintegralfs}) is given by $-2eL(aH)^3$ times
\begin{equation}\label{Lambda0thcontribtion2}\begin{split}
& -\frac{\gamma_E}{6\pi^2}+\frac{37}{144\pi^2}-\frac{L^2}{120\pi^2}-\frac{L^4}{420\pi^2}+\frac{7M^2}{192\pi^2} +\frac{3M^2}{8\pi^2L^2}
-\frac{L^2M^2}{1440\pi^2}+\frac{5M^4}{576\pi^2} + \frac{3M^2 r}{16\pi^2L^3}\log\left(\frac{r-L}{r+L}\right)\\
&  -i\big(\mathrm{imaginary\; part\; of\; (\ref{Lambda0thcontribtion1})}\big) 
- \frac{r}{48\pi^5 L^2 \sinh(2\pi r)}\Bigl\{(45-\pi^2(11-12L^2+8r^2))\cosh(2\pi L)\\ 
&\qquad\qquad\qquad\qquad\qquad\qquad\qquad\qquad\qquad\qquad\qquad\qquad  - (45-\pi^2(11-72L^2+8r^2))\frac{\sinh(2\pi L)}{2\pi L}\Bigl\}  \\ 
& + \frac{3rM^2}{32\pi^2 L^3\sinh(2\pi r)}\sum_{s=\pm} s\mathrm e^{2\pi rs}(\mathrm{Ei}(2\pi s(r+L)) - \mathrm{Ei}(2\pi s(r-L))) \\
& -\Re[\int_{-1}^1 \mathrm dx \frac{(1+r^2-(1+3L^2+3r^2)x^2+5L^2x^4)}{16\pi^2 \sinh(2\pi r)} \\
& \qquad\qquad\qquad \times ((\mathrm e^{2\pi Lx}-\mathrm e^{2\pi r})\psi(i(Lx+r)) - (\mathrm e^{2\pi Lx}-\mathrm e^{-2\pi r})\psi(i(Lx-r)) ) ]. 
\end{split}
\end{equation}
Finally, Eqs.~(\ref{divpart}), (\ref{Lambda0thcontribtion1}) and (\ref{Lambda0thcontribtion2}) yield Eq.~(\ref{formalJ3}).

\bibliographystyle{JHEP.bst}
\bibliography{ref_fermionicschwinger}

\end{document}